\documentclass[a4paper,10pt,oneside,onecolumn,preprint,authoryear]{elsarticle}
\usepackage[utf8]{inputenc}

\usepackage{graphicx}
\usepackage{setspace}
\usepackage{hyperref}
\begin{document}
\title{Reconstructing the size distribution of the primordial Main Belt}

\author[aob,oca]{G.~Tsirvoulis\corref{cor1}}
\ead{gtsirvoulis@aob.rs}

\author[oca]{A.~Morbidelli}
\ead{morby@oca.eu}

\author[oca]{M.~Delbo}
\ead{marco.delbo@oca.eu}

\author[auth]{K.~Tsiganis}
\ead{tsiganis@auth.gr}

\address[aob]{Astronomical Observatory, Volgina 7, 11060 Belgrade 38, Serbia}
\address[oca]{Observatoire de la Côte d'Azur, Boulevard de l’Observatoire, 06304 Nice, France}
\address[auth]{Department of Physics, Aristotle University of Thessaloniki, 54124 Thessaloniki, Greece}

\cortext[cor1]{Corresponding author}

\begin{abstract}
In this work we aim to constrain the slope of the size distribution of main-belt asteroids, at their primordial state. To do so we turn out attention to the part of the main asteroid belt between 2.82 and 2.96~AU, the so-called ``pristine zone'', which has a low number density of asteroids and few, well separated asteroid families. Exploiting these unique characteristics, and using a modified version of the hierarchical clustering method we are able to remove the majority of asteroid family members from the region. The remaining, background asteroids should be of primordial origin, as the strong 5/2 and 7/3 mean-motion resonances with Jupiter inhibit transfer of asteroids to and from the neighboring regions. The size-frequency distribution of asteroids in the size range $17<D(\rm{km})<70$ has a slope $q\simeq-1$. Using  Monte-Carlo methods, we are able to simulate, and compensate for the collisional and dynamical evolution of the asteroid population, and get an upper bound for its size distribution  slope $q=-1.43$. In addition, applying the same 'family extraction' method to  the neighboring regions, i.e. the middle and outer belts, and comparing the size distributions of the respective background populations, we find statistical evidence that no large asteroid families of primordial origin had formed in the middle or pristine zones. 
\end{abstract}
\maketitle
\section{Introduction}
One of the reasons for which asteroids are subject of many studies is that they represent what is left over of the original population of planetesimals in the inner Solar System. Among the many properties of asteroids, their size-frequency distribution\footnote{The size-frequency distribution of asteroids is usually approximated by a power law, with a characteristic exponent q: $N(D)\sim D^q$}  (SFD) may be diagnostic of the processes by which planetesimals formed. The current cumulative SFD of asteroids is characterized by a quite steep slope in the size range $100\rm{km}<D<1,000\rm{km}$ (with an exponent q of about $-2.5$), and a shallower slope for $D<100\rm{km}$ (with $q\sim-1.8$ down to $D\sim10\rm{km}$). The current SFD of the asteroids, however, is presumably not identical to the SFD that planetesimals had at the time of their formation, but has evolved over the age of the Solar System as a consequence of various phenomena: collisions between asteroids produce a plethora of small fragments from only two original bodies, directly altering the SFD of the total population. Moreover, dynamical depletion is constantly removing asteroids from the main belt: The interplay between the Yarkovsky thermal force and the strongest resonances (mean-motion and secular ones), is the most important depletion mechanism, and since the Yarkovsky effect is size-dependent the SFD is modified accordingly.

Using several observational constraints, \cite{Bottke2005} concluded that the original SFD of planetesimals below $D=100\rm{km}$ had to be equal to or shallower than the current one. However, they could not constrain what the original slope had to be. Considering the possibility of a very shallow primordial slope, \cite{Morbidelli2009} suggested that asteroids formed big, with characteristic sizes in the 100~km-1,000~km range. The model emerging at the time about planetesimal formation from massive self-gravitating clumps of dust (\cite{Cuzzi2008}) and pebbles (\cite{Johansen2007a}) seemed to support, at least qualitatively, that claim.

More recently, \cite{Johansen2015} studied in details the formation of planetesimals by streaming instability \citep{Youdin2007,Johansen2007b}, using hydrodynamical simulations with multiple resolutions. They found that the planetesimals formed by this process have a characteristic cumulative SFD with exponent $q=-1.8$. Because this slope is very close to that currently observed for asteroids with $D<100\rm{km}$, \citet{Johansen2015} proposed that 100~km is the maximal size of the planetesimals formed by the streaming instability. The asteroids currently larger than 100~km would have grown from primordial sizes smaller than this upper limit in a subsequent process named "pebble accretion" (\cite{Lambrechts2012}).  \cite{Klahr2016} instead, found that the characteristic size of the planetesimals formed by the streaming instability is $D\sim100\rm{km}$, but with the wings of the size probability function extending to smaller and larger bodies. \cite{Cuzzi2010} had obtained a similar result, but for the turbulent concentration of clumps of small particles, rather than the streaming instability.

It is clear that the papers quoted above about planetesimal accretion, present quite different views on the characteristic sizes of the first planetesimals. In order to discriminate among them, it would be important to have an observational assessment of the primordial planetesimal SFD below 100~km in size. But, as said above, the asteroid SFD has evolved through collisions and dynamical depletion.

In principle, asteroids (tens of km in diameter) produced in collisions should be identifiable as members of asteroid families. Thus, if  one removes the asteroid families from  asteroid catalogs, one should be left with the population of these bodies which have not been produced by collisions through the lifetime of the Solar System: namely the primordial population. However, this procedure is not so easy to implement. The Hierarchical Clustering Method (\cite{Zappala1990}) the most used procedure for the identification of asteroid family members  usually succeeds in linking only the compact core of the family. This has been shown by \cite{Parker2008}, who demonstrated that each nominal family identified by HCM is surrounded by a halo sharing the same spectral properties. Recent upgrades of the HCM (\cite{Milani2014}; \cite{Milani2016}) attempt to identify the family halos through a multi-step approach. However, it is unlikely that the entire family population can be identified with confidence even with this more sophisticated approach. The situation may be better for relatively large asteroids that we are interested here, but this is not certain.

Here, we assess the fraction of the background asteroid population (i.e. the population not
belonging to any family) that is made of rogue family members and the characteristic size
at which this contamination starts to be relevant. To do so, we focus in a zone of the
asteroid belt, with semi-major axis $2.82< a < 2.96$ AU, which contains much fewer
asteroids than any other zone. The explanation for this deficit of asteroids, according to
\cite{Broz2013}, is due to the bordering of the 5/2 and the 7/3 mean motion resonances
with Jupiter, which prevent the influx of asteroids migrating due to the Yarkovsky effect
\citep{Bottke2002} from the neighboring regions. Also, because the region is quite narrow, only few
asteroid families formed in it. For these reasons, \cite{Broz2013} dubbed this region as
the ”pristine zone”, as it is probably the one that reflects the best the primordial
distribution of asteroids.

 In this region it is fairly easy to subtract the family members, given the small number of families and the low orbital density of the overall population.  We can also try to subtract all family members from the two regions that border the pristine zone, which contain a larger number of asteroids. This procedure is explained in section 2. In principle, there is no reason that the primordial orbital densities of asteroids were different in neighboring regions. Thus, in Section 3, by comparing the nominal background population in the pristine zone with those in neighboring regions with the same semi major axis width, we can get statistical information on which fraction of these neighboring background populations should be in reality made of rogue family members that we cannot identify as such.  

We then go further in our analysis in Section 4. To gain confidence that the background population in the pristine zone really represents the primordial SFD of asteroids and to determine up to which absolute magnitude this is true, we compare it with those in the neighboring regions. We require that at least in one of the neighboring zones the SFD of the background population is the same as in the pristine zone (e.g. same shape, same slopes, number of asteroids within a factor of $\sim 2$). We find that this is the case in the inner neighboring region up to absolute magnitude $\rm{H}\sim12$, while we explain why the outer neighboring zone is different. Moreover we verify that the background SFD for $H<12$ in the  pristine zone is different from those of the families in these two regions, as suggested by \citet{Cellino1991}.

Based on these results, in Sect. 5 we measure the slope of the SFD of the background population in the pristine zone between $9<H<12$. However, this is not yet the slope of the SFD of the primordial planetesimals below 100~km in size, because some original asteroids in this magnitude range might have been destroyed by collisions, even if, in principle, none of the current background asteroids was produced by collisions (by definition of background, if selected correctly). Thus, we correct the SFD slope by the size-dependent probability to have been catastrophically disrupted over the age of the Solar System, given in \cite{Bottke2005}. Finally, we compare this slope with that expected by the streaming instability in the \cite{Johansen2015} simulations.

The conclusions of this work are summarized and discussed in Sect. 6.

\section{Identification of family members}

\begin{figure}[h!]
\begin{center}
\includegraphics[width=\textwidth]{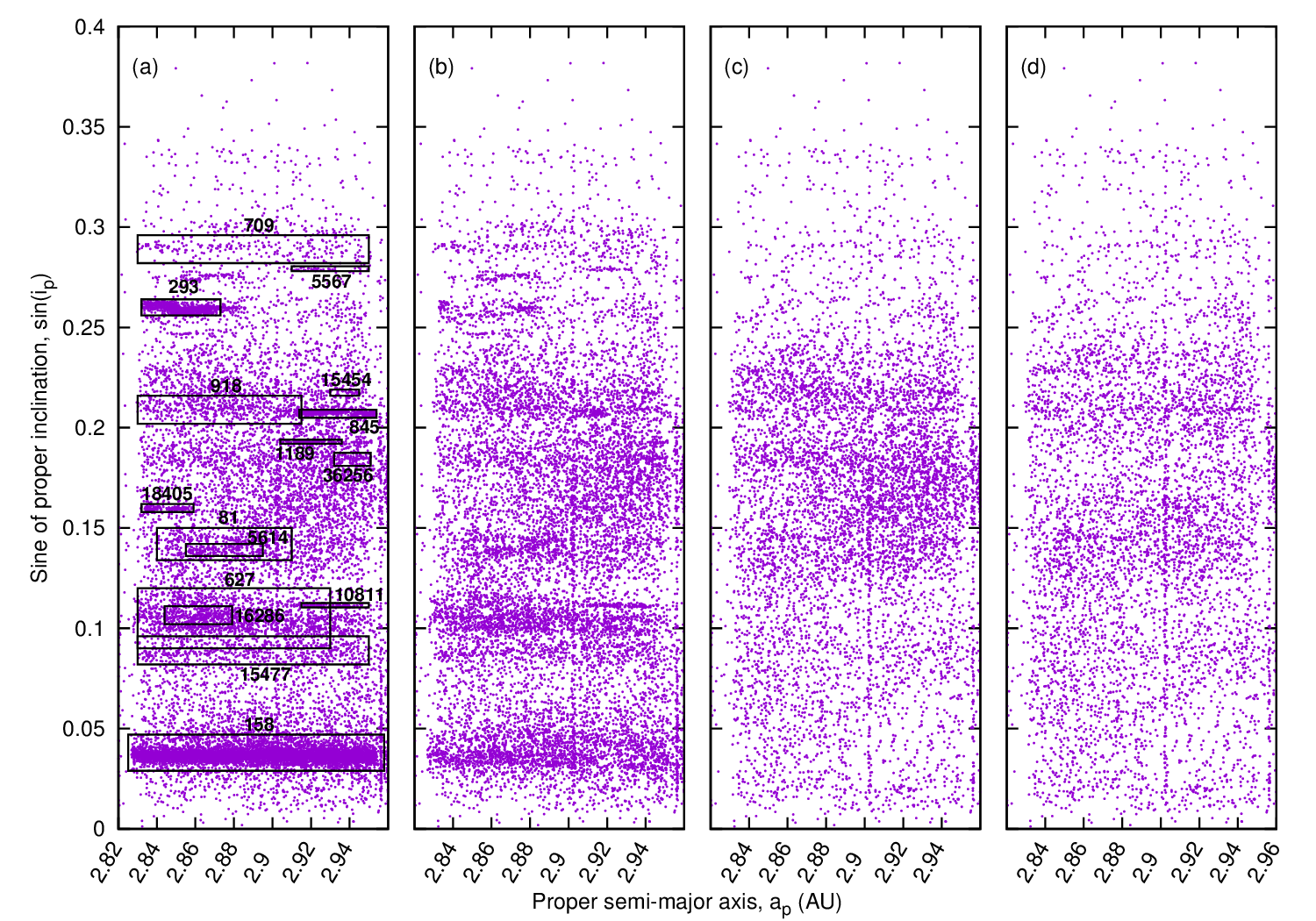} 
 \caption{Asteroids in the pristine zone of the main belt in the proper semi-major axis versus sine of proper inclination plane. Panel a: all numbered and multi-opposition asteroids. The boxes highlight the asteroid families in the region. Panel b: The remaining asteroids after removing the family members according to the classifications of \cite{Milani2014}, \cite{Milani2016} and \cite{Broz2013}. Panel c: The remaining asteroids after removing family members with our method as discussed in the text. Panel d: Same as panel c, with the asteroids originating from the family of Eos also removed.}
   \label{Fig:pri3}
\end{center}
\end{figure}

The first step of our study is to obtain the background population of the pristine zone. To do so we simply remove from the catalog of proper elements of numbered and multi-opposition asteroids \footnote{Obtained from: http://hamilton.dm.unipi.it/astdys/index.php?pc=5} those asteroids that have been identified as family members following the classification of \cite{Milani2014} and \cite{Milani2016}. However due to the fact that the focus of their study was to obtain a good classification of families, the authors of these works adopted a conservative approach in the selection of their Quasi Random Level (QRL)\footnote{The Quasi Random Level is a measure of the statistical significance  when identifying asteroid families. It sets a threshold on the cut-off velocity, the maximum distance in the proper elements space between asteroids belonging to the same group, above which there is no statistical difference between an actual family and a statistical fluke of a random distribution of asteroids. For more see: \cite{Zappala1990}}for the hierarchical clustering analysis \cite{Zappala1990}, in order to avoid background objects from being incorrectly identified as family members and maintain good separation in orbital elements between families. Moreover they used the same QRL parameter for the pristine zone as for the rest of the outer belt. This resulted in a statistically significant family identification, which however left as background a lot of asteroids that should belong to the halos of asteroid families. This can be appreciated by looking at \autoref{Fig:pri3} panel b, where we see that even after removing all family members according to the Milani et al. classification, most of the very same families are still recognizable by the density contrast in the proper element space. For our purpose, which is to obtain as clear of a background as possible this is not the optimal solution. Therefore we decided to proceed with a modified application of the hierarchical clustering method, trying to get rid of as many family members as possible.
We perform the hierarchical clustering method to the catalog of proper elements of the pristine zone, starting with the parent bodies identified by \cite{Milani2014}. Moreover we also consider the parent bodies of asteroid families identified in \cite{Broz2013} which are not present in the \cite{Milani2014} classification, to make sure we remove as many family members as possible. We obtain for each asteroid family the number of associated members as a function of the cut-off velocity. We vary the latter, in increments of 2 m/s, from very small values where no close neighbor is found, up to the point where the family includes an abnormally large portion of the total population of asteroids. Then we select for each asteroid family the optimal cut-off velocity in the following way: We select the highest possible value at which each family is still identifiable as a single cluster of asteroids, before merging with the background. If two families are merged together at some value of the cut-off velocity, we consider them as a group and go on until the background is included, assigning to the group the cut-off velocity value of the previous step before merging with the background happens. 

\begin{figure}[h!]
\begin{center}
\includegraphics[width=\textwidth]{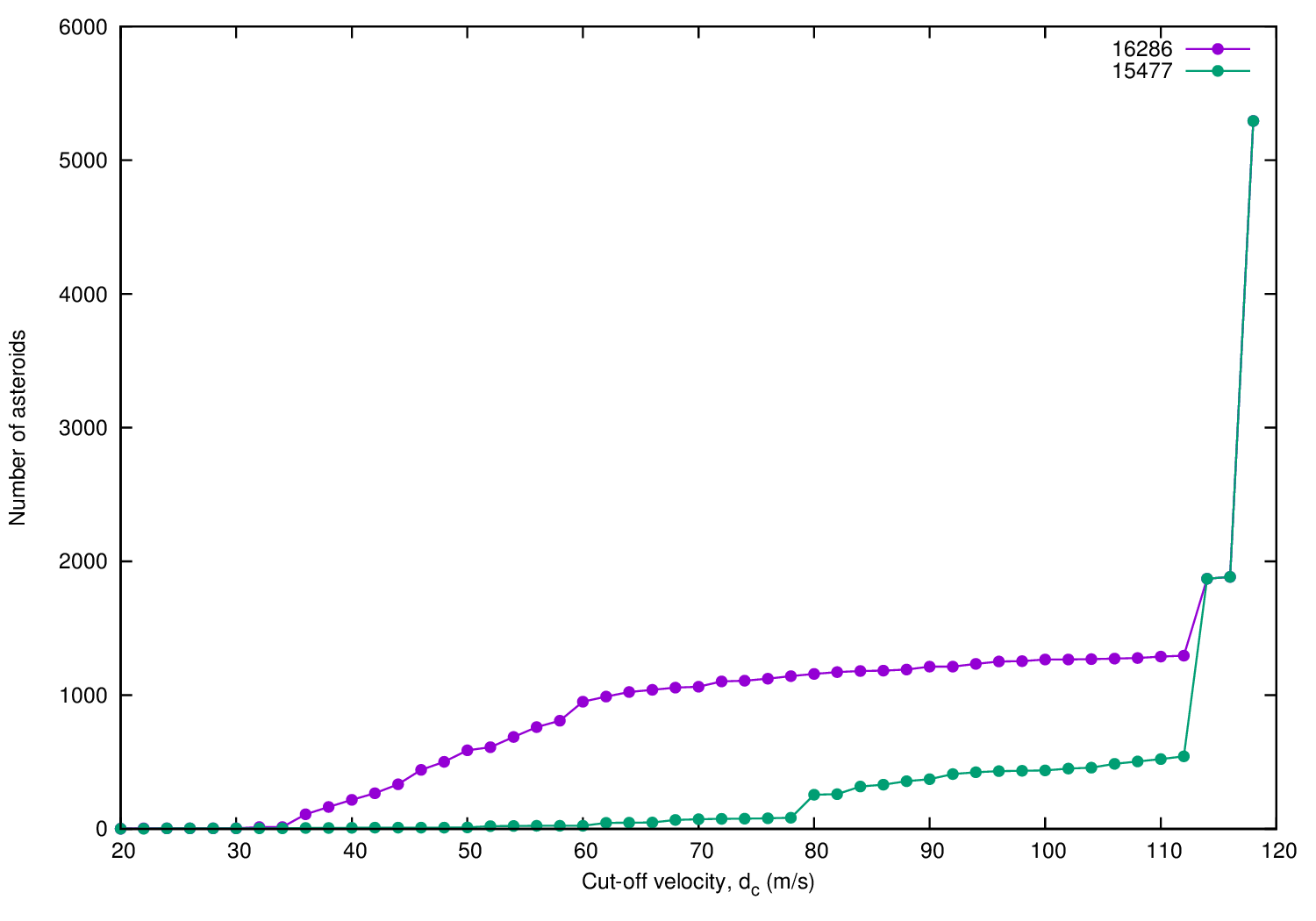} 
 \caption{The number of asteroids associated to asteroids (16286) and (15477) as a function of cut-off velocity.}
   \label{Fig:16286}
\end{center}
\end{figure}

As an example, the family of (16286) was found by \cite{Milani2014} to have 83 family members at a cutoff velocity of 40 m/s. Our method gives the result seen in \autoref{Fig:16286}. The family membership starts growing linearly with increasing cut-off velocity from 34 m/s up to 60 m/s. Then between 60 m/s and 112 m/s it grows at a much smaller rate, giving the distinctive ``plateau'', the midpoint of which is often used as the nominal cut-off when aiming for a reliable family membership (see e.g. \cite{Novakovic2011}). At 114 m/s it merges with the family of (15447) (identified by \cite{Broz2013}), and they both merge with the background at 118 m/s. The family of (15477) was found by \cite{Broz2013} to have 144 members at 110 m/s. In this case since the two families do not grow as a group after they merge together until they extend to the background, we select the value of 112 m/s for both families resulting in 1296 members for (16286) and 542 members for (15477). Note that these membership numbers are considerably larger than the respective ones given by the aforementioned authors. In the case of (16286) although we use the same catalog of proper elements both the identification method and the selection of cut-off velocity are different. We use a straightforward application of the HCM compared to the multi-step procedure of \cite{Milani2014}, and we select on purpose a very high cut-off velocity compared to the QRL approach of the latter. In the case of (15477) although the methods and the cut-off velocity are the same, the catalog of proper elements used is different, since we use a more updated version which also contains multi-opposition asteroids.

In all cases we end up with more family members than in the works of other authors. We know that many of these asteroids which we identify as family members are in reality interlopers, and thus we do not claim to have produced another family classification. Our aim was to obtain a background population contaminated as little as possible by family members. In doing so we lose a lot of background asteroids into the families, but we expect that the number of background objects lost in this way is too small to have a significant effect on the resulting size distribution. The final result is seen in \autoref{Fig:pri3} (c), where we see that we have removed substantially more family members compared to (b). The background we obtain is much more uniform. 

\begin{figure}[h!]
\begin{center}
\includegraphics[width=\textwidth]{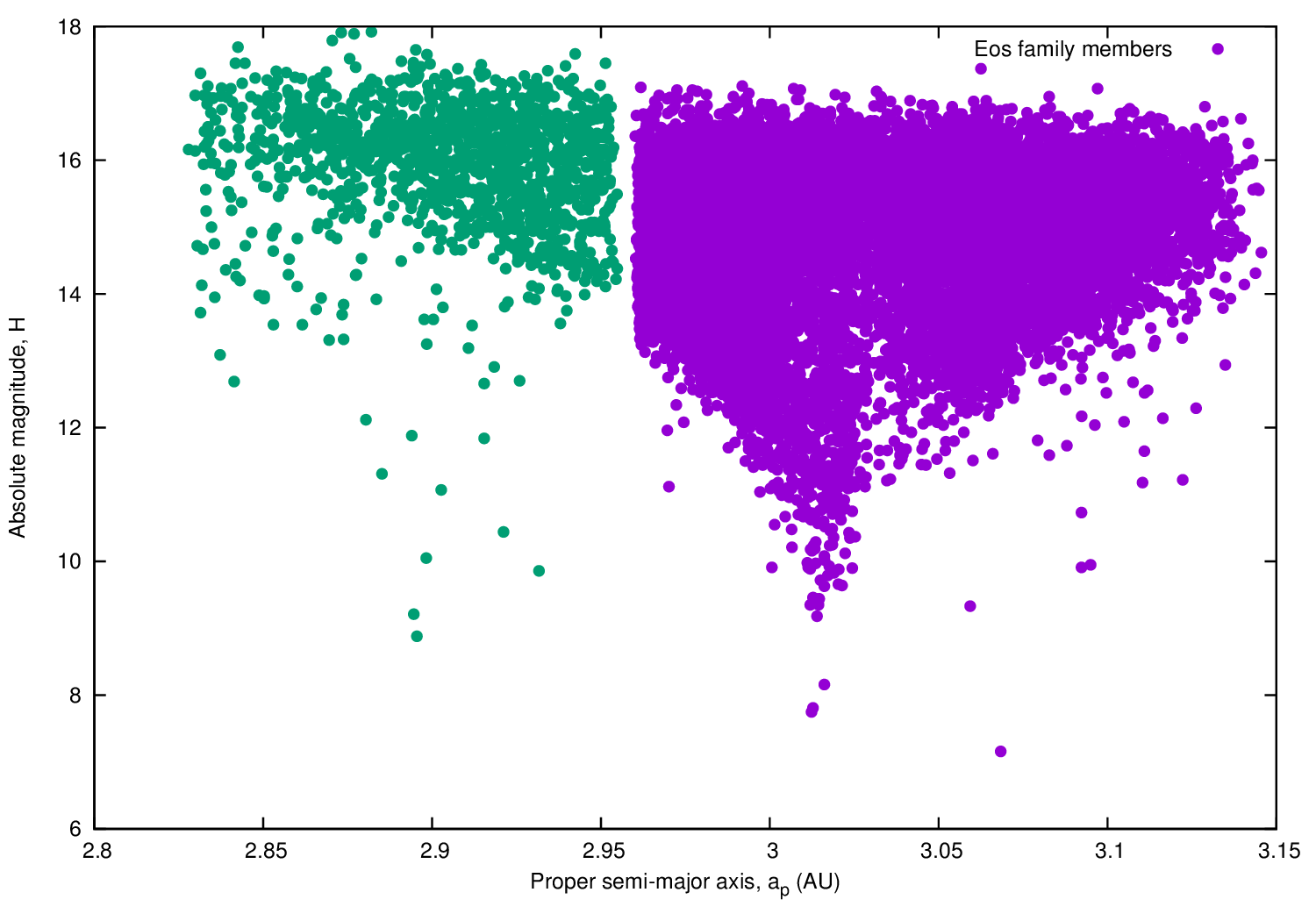} 
 \caption{The asteroid family of (221) Eos in the proper semi-major axis versus absolute magnitude plane. In purple are the asteroids belonging to the classical family, as identified by \cite{Milani2014}, while in green are asteroids in the pristine zone with $2.82<a_{\rm{p}}<2.96$, $0.03<e_{\rm{p}}<0.1$ and $0.12<\sin{i_{\rm{p}}}<0.2$. Notice that the V-shape of Eos extends into this population.}
   \label{Fig:eos}
\end{center}
\end{figure}

Still, in \autoref{Fig:pri3} (c), there is one extended concentration of asteroids in the range $0.12<\sin{i_{\rm{p}}}<0.25$ which needs to be further investigated. For the asteroid families of (1189), (16286) and (36256), this population of asteroids is included in their membership list at a cut-off velocity one step higher than the one selected. This means that the maximal cut-off velocity we used for these three small families was in fact chosen in order to avoid merging them together with this large concentration of asteroids. However, visual inspection of \autoref{Fig:pri3} (c) suggests that this concentration is not the halo of either of the mentioned families but instead it is an independent family, previously unidentified. If this is true, this concentration should have some properties distinctive of families. One such property is the so-called ``V-shape'', that is the shape of the distribution in the $(a_{\rm{p}},H)$ plane that a family has to acquire due to the size-dependent action of the so-called Yarkovsky effect. We selected all background asteroids from the previous step in the volume containing this concentration, i.e. $2.82<a_{\rm{p}}<2.96$, $0.03<e_{\rm{p}}<0.1$ and $0.12<\sin{i_{\rm{p}}}<0.2$, and plotted them in the $(a_{\rm{p}},H)$ plane as seen in \autoref{Fig:eos}. The result is striking. The left half of a V-shape is clearly visible, meaning that the other half must exist at larger semi-major axes. But this range in eccentricities and inclinations matches almost perfectly the range covered by the family of (221) Eos in the outer belt. Indeed plotting the family members of (221) Eos in the same plane shows that it extends into the pristine zone, creating this mysterious high concentration of asteroids. This is proving that this region is not so pristine as previously believed (\cite{Broz2013}), and it can indeed be contaminated by asteroids drifting from the adjacent regions. To remove those asteroids within this orbital volume originating from Eos, we took a step back, and used a higher cut-off value for the neighboring family of (1189). In this way we remove the families of (1189), (16286) and (36256), together with asteroids coming from Eos in one step.

\begin{figure}[h!]
\begin{center}
\includegraphics[width=\textwidth]{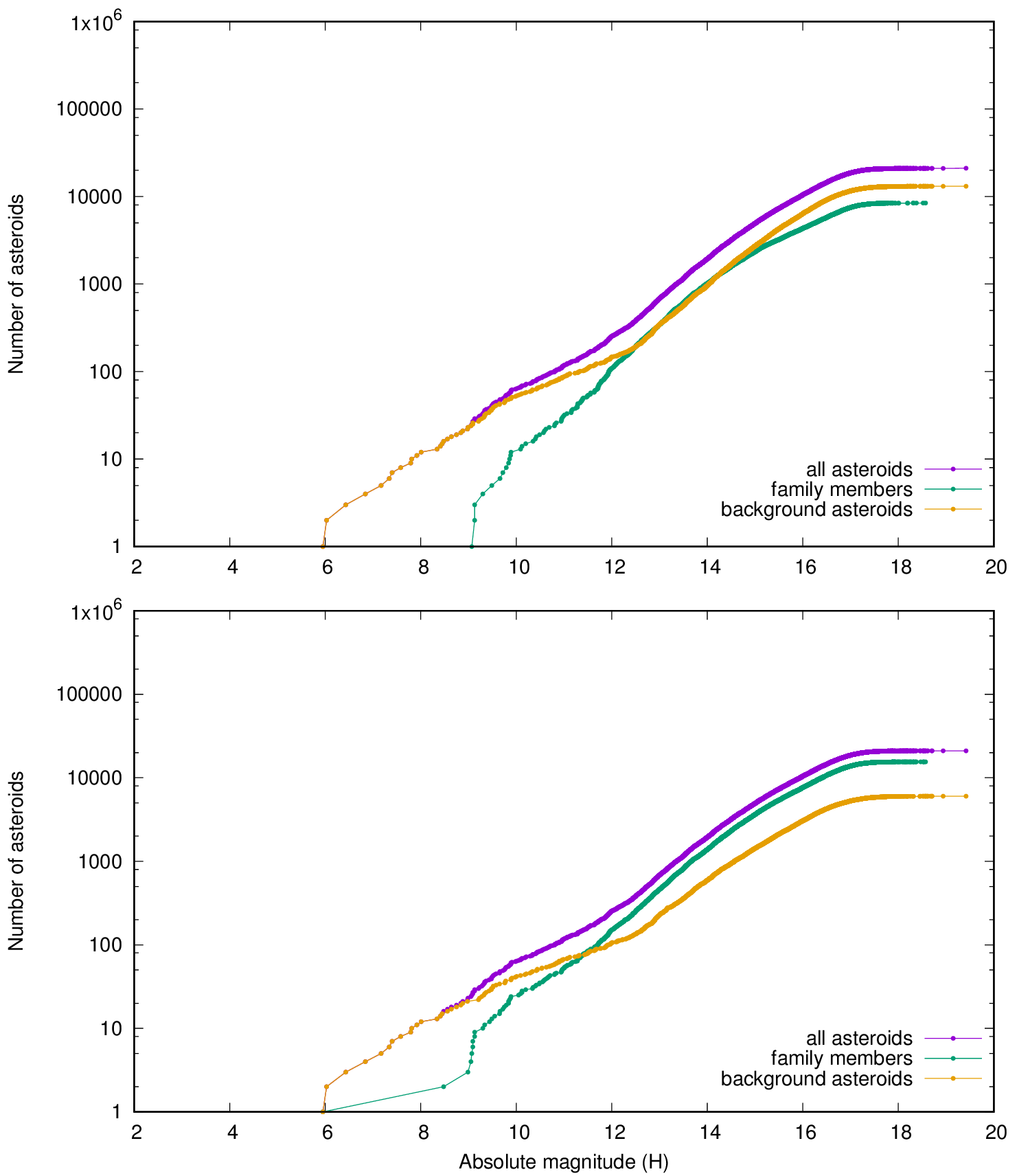} 
 \caption{Cumulative size distribution of asteroids in the pristine zone according to the classification of \cite{Milani2014} (top) and the one in this work (bottom). The colors represent: all asteroids (purple), asteroid family members (green) and background objects (orange).}
   \label{Fig:pristine}
\end{center}
\end{figure}

The final result is shown in \autoref{Fig:pri3} (d). \autoref{Table:1} shows a summary of the numbers of family members and background asteroids at each step, and \autoref{Fig:pristine} presents the corresponding size distributions.

\begin{table}[]
\centering
\caption{Number of asteroids classified as family members and background objects according to \cite{Milani2016} and our method.}
\label{Table:1}
\begin{tabular}{|l|c|c|c|l|}
\hline
 & \multicolumn{1}{l|}{Milani et al. 2016} & \multicolumn{1}{l|}{Extended families} & \multicolumn{2}{l|}{\begin{tabular}[c]{@{}l@{}}Extended families \\ + Eos members\end{tabular}} \\ \hline
Family members & 8430 & 13839 & \multicolumn{2}{c|}{15533} \\ \hline
Background & 13122 & 7713 & \multicolumn{2}{c|}{6019} \\ \hline
Total & \multicolumn{4}{c|}{21552} \\ \hline
\end{tabular}
\end{table}

\begin{figure}[h!]
\begin{center}
\includegraphics[width=\textwidth]{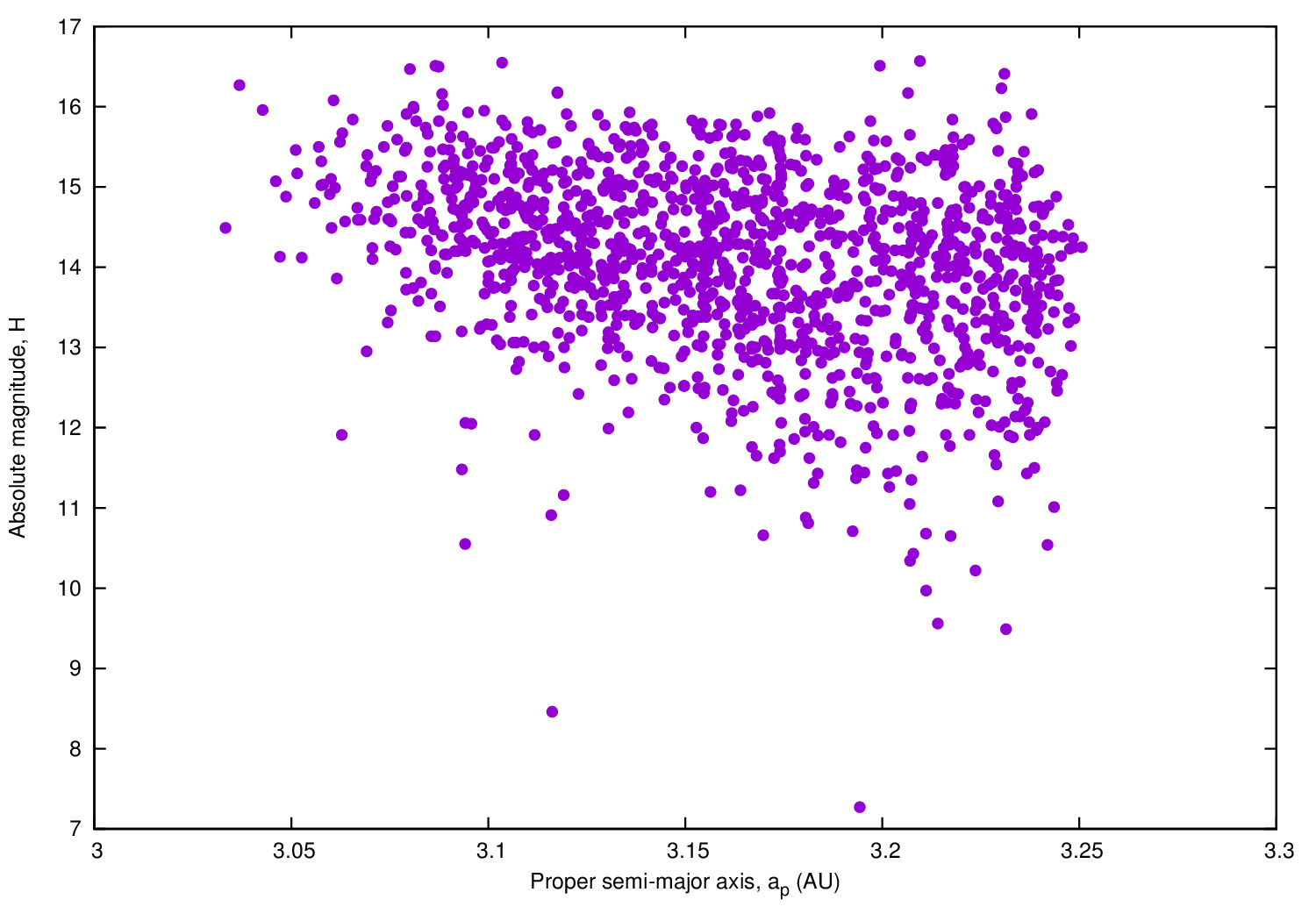} 
 \caption{The distribution of asteroids of the family around (892) Seeligeria on the $(a_{\rm{p}},\rm{H})$ plane. Note the apparent left half of a V-shape.}
   \label{Fig:6355}
\end{center}
\end{figure}

\begin{figure}[h!]
\begin{center}
\includegraphics[width=\textwidth]{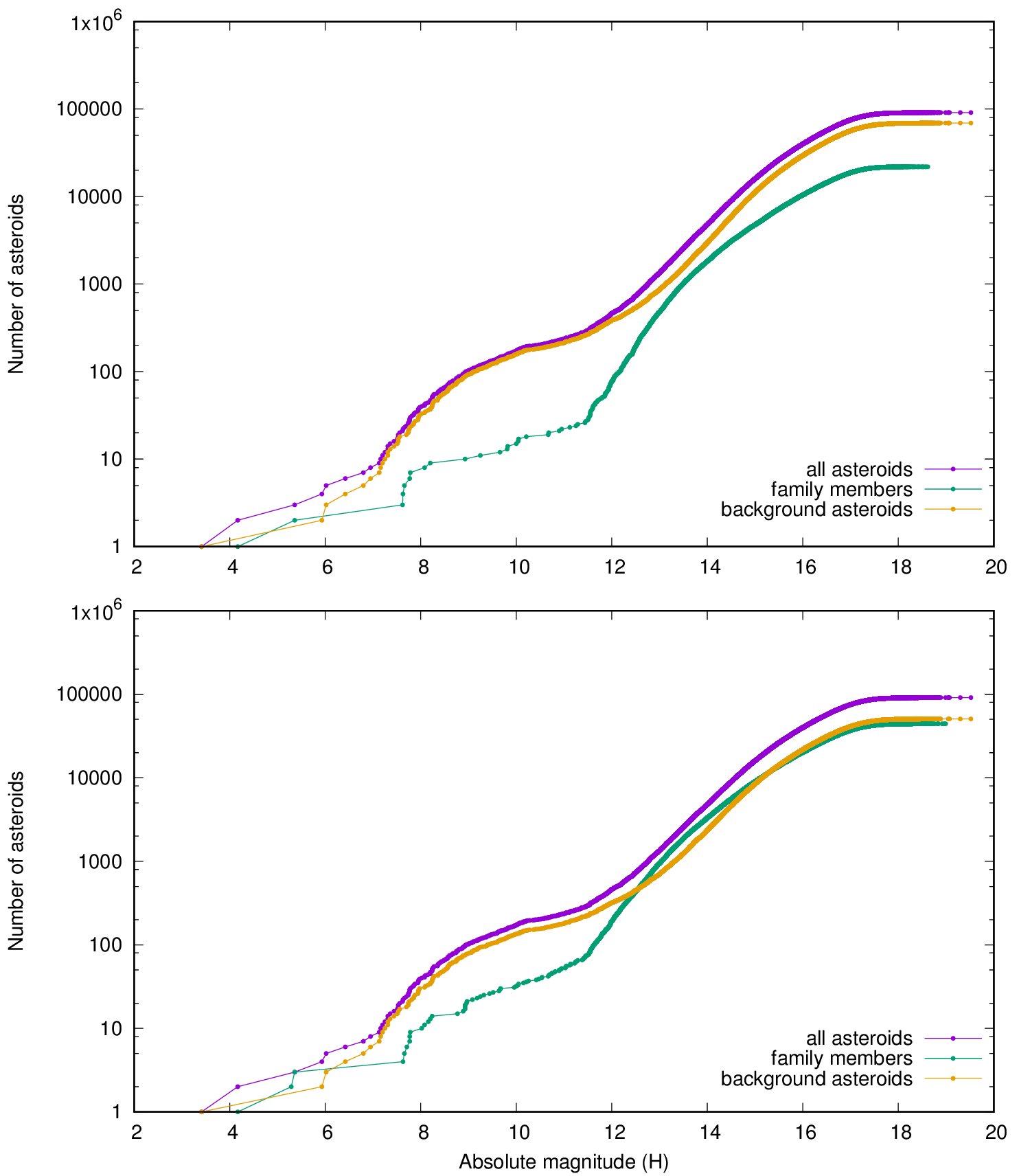} 
 \caption{Same as \autoref{Fig:pristine} for the middle belt}
   \label{Fig:mid}
\end{center}
\end{figure}

\begin{figure}[h!]
\begin{center}
\includegraphics[width=\textwidth]{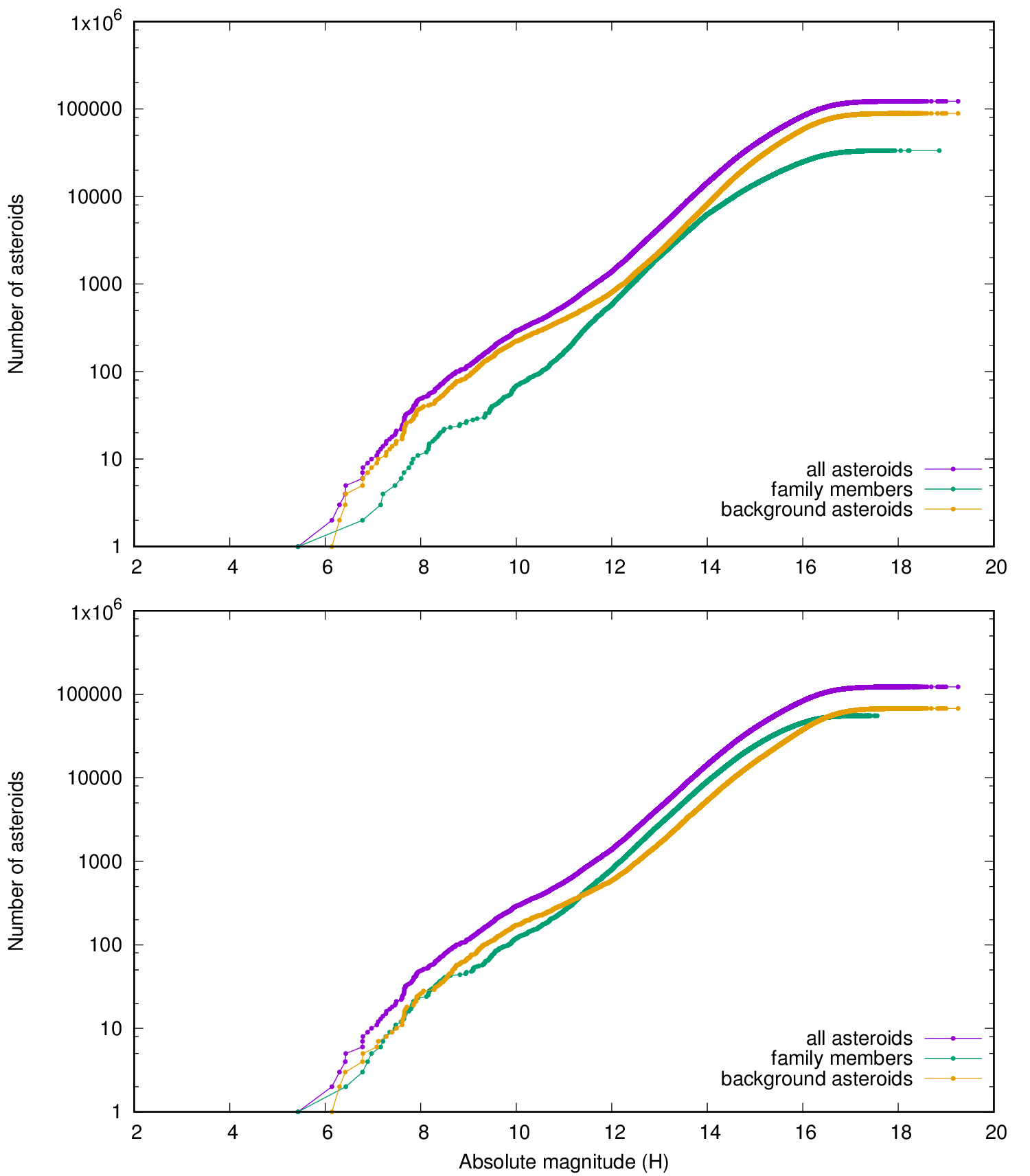} 
 \caption{Same as \autoref{Fig:pristine} for the outer belt}
   \label{Fig:out}
\end{center}
\end{figure}

We then performed the same procedure of removing family members from the middle and outer belts\footnote{In some works the outer belt is considered to extend from the 5/2 out to the 2/1 MMRs with Jupiter ($2.82<a_{\rm{p}}<3.26$~AU). Since in this work we treat the pristine zone separately, we use as the limits of the outer belt the 7/3 and 2/1 MMRs with Jupiter ($2.96<a_{\rm{p}}<3.26$~AU)} . For the middle belt we chose to analyze the region with $2.65<a_{\rm{p}}<2.82$, excluding its innermost part. This choice was made because that part contains the asteroid family of (5) Astraea, which is very disperse and has a large halo, and is crossed by several secular resonances.This makes our identification method useless given that almost all asteroids form a large clump with a small increase of the cut-off velocity. As the middle and outer belts have a much higher number density of asteroids and many more asteroid families than the pristine zone, the application of our method of extending the family membership was more challenging. More families had to be treated together as groups due to their proximity, and the choice of cut-off velocity for each case was not so straightforward. For example, in order to remove asteroids belonging to the family of (221) Eos, as we increase the cut-off velocity the families of (179), (283), (507), (8737) and (21885) were merged with (221) resulting in a big cluster of $\sim35000$ asteroids. Also in the outer belt, in the range $0.35<\sin{i_{\rm{p}}}<0.42$, \cite{Milani2014} identified four small families, namely (1101), (3025), (6355) and (10654), whereas by increasing the cut-off velocity we find that almost the whole region merges into one large family. We argue that this new big family is real, based on the apparent half ``V-shape'' of its members on the $(a_{\rm{p}},\rm{H})$ plane (see \autoref{Fig:6355}), a characteristic of asteroid families and not of random samples of asteroids. The size distributions of the family members and background asteroids for the middle and outer belts are shown in \autoref{Fig:mid} and \autoref{Fig:out} respectively.

\section{Background asteroids vs. rogue family members}

\begin{figure}[h!]
\begin{center}
\includegraphics[width=\textwidth]{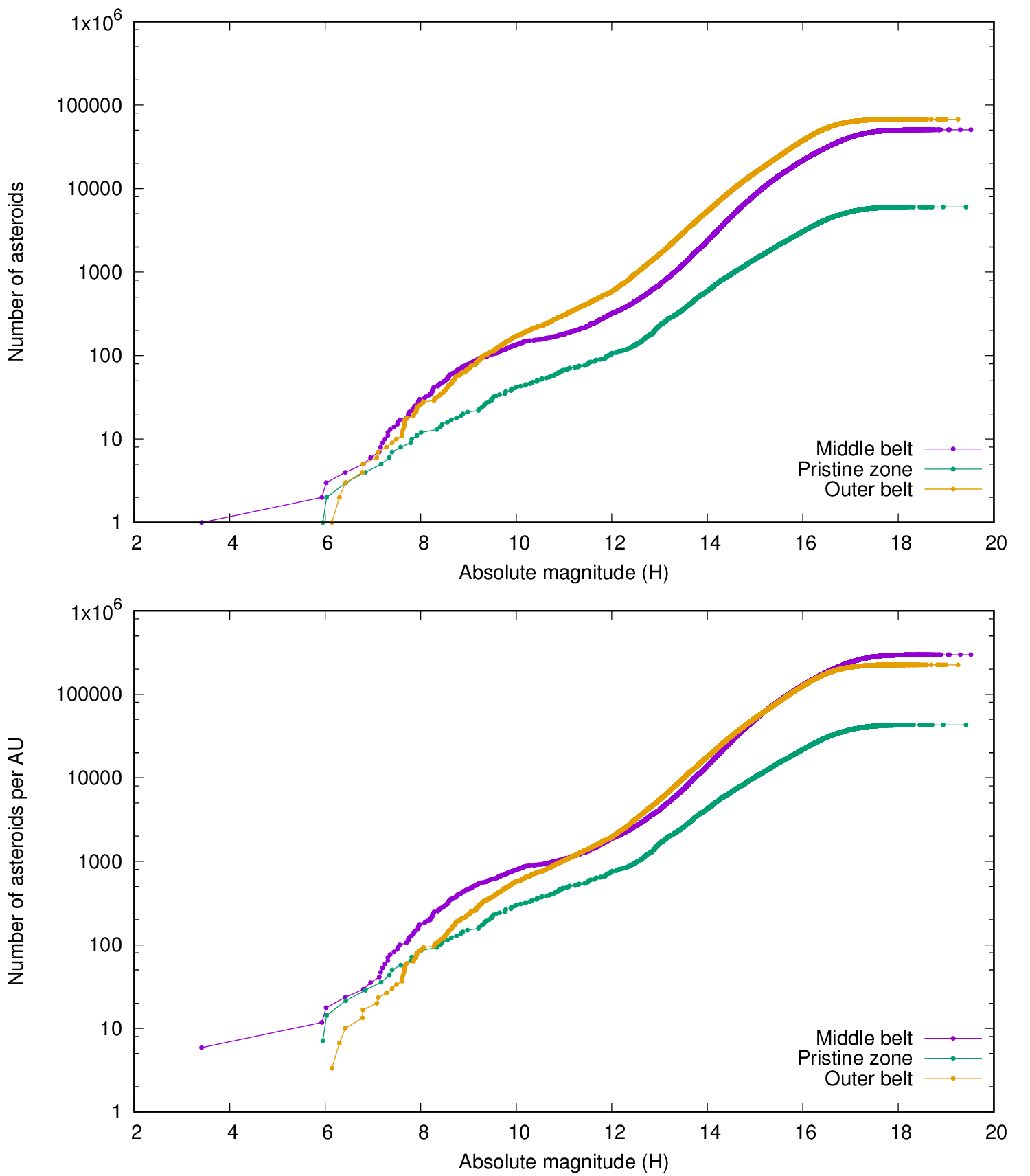} 
 \caption{The cumulative absolute magnitude distribution of background asteroids in the three regions. In the bottom panel the population of each region is normalized in terms of semi-major axis.}
   \label{Fig:bg}
\end{center}
\end{figure}

The first piece of statistical information we can extract from the size distributions obtained, is which fraction of the background population is made of primordial asteroids and which of collisionally generated ones. To do this we turn our attention to the size distributions of background objects in the three regions as shown in \autoref{Fig:bg}. To compare the three regions we study here, we normalized the populations of asteroids in the three regions in terms of the orbital volume they contain, essentially dividing the number of asteroids in each region with its corresponding semi-major axis range. In the bottom panel of \autoref{Fig:bg} we show the absolute magnitude distributions of background asteroids per AU for the middle belt, the pristine zone and the outer belt. The first thing we notice is that the total number of asteroids in the middle and outer belts is substantially larger than that in the pristine zone. This result was expected, as in the more populous regions of the main belt we can't remove all family members by applying  the HCM, but the consequence is rather surprising: More than 80\% of what we would consider as the ``background'' population of the middle and outer belts is in fact rogue family members, as the number of asteroids per astronomical unit in these regions is  about seven times larger than in the pristine zone. \citet{Cellino1996} predicted, based on the difference in the slopes of the SFDs of family members and background asteroids, that more than 90\% of discovered small asteroids should belong to asteroid families. Our result not only verifies, but also reinforces their prediction, as we find that even our aggressively obtained background consists in fact mainly of collisionally generated asteroids. This means that the vast majority of the asteroids  we currently observe in the main belt, even when they are not identified as family members, are products of collisional evolution, rather than primitive bodies.

\section{The collisional history of the main-belt}

\autoref{Fig:bg} reveals two key aspects regarding our study of the primordial distribution of main-belt asteroids: The first is that the magnitude distributions of the background population in the middle and pristine zones share the same qualitative characteristics, especially in the range $9<H<12$ of absolute magnitudes, as their slopes  in this range are $q_{mid}\simeq4.8$ for the middle belt and $q_{pri}\simeq5.2$ for the pristine zone. The second, obviously but equally important aspect, is that they differ from the size distribution of the outer belt, as the latter has a slope of $q_{out}\simeq7$ in the same range of absolute magnitudes.

The first aspect is a strong suggestion that the background population of the pristine zone we have obtained should reflect the size distribution of primordial asteroids with $H<12$. This claim is also supported by the fact that the composite magnitude distribution of family members in the pristine zone differs drastically from the one of the background (\autoref{Fig:pristine}). This confirms that the families in the pristine zone have been adequately removed, and do not contaminate the background significantly.

The second aspect actually concerns not only the background populations, but also the SFD of all families together, as evident when comparing \autoref{Fig:mid} and \autoref{Fig:out}, where we see that not only the backgrounds but also the composite populations of family members in the three regions have different magnitude distributions. As we explained above, the asteroids originating from family creating events dominate the populations in the middle and outer belts, and as a consequence the background size distributions in each region. This means that the difference in the shape of the size distributions of the middle and outer belts should reflect different collisional records. Thus, we seek the cause of the difference in the distributions of the populations of the middle and outer belts by looking into the individual families therein. According to \citet{Durda2007} the conditions and scale of the family forming event (e.g. cratering vs. catastrophic event\footnote{Asteroid families are usually classified as being of the cratering type, if the volume of the largest remnant is much larger than the sum of the volumes of the rest of the family members ($>90\%$), suggesting that the family was formed from material excavated from a relatively small crater on the parent body. If, on the other hand, the volume of the largest remnant is comparable to the sum of the volumes of the other family members ($<90\%$), the asteroid family is classified as being of the catastrophic or fragmentation type. In this case the family-forming impact was severe enough to completely fragment the parent body. The value of $90\%$ used for the ratio of the volumes to distinguish between the two types is used conventionally.}, impactor velocity and incident angle, size ratio etc.) are reflected on the SFD of family members.

\begin{figure}[h!b]
\begin{center}
\includegraphics[width=\textwidth]{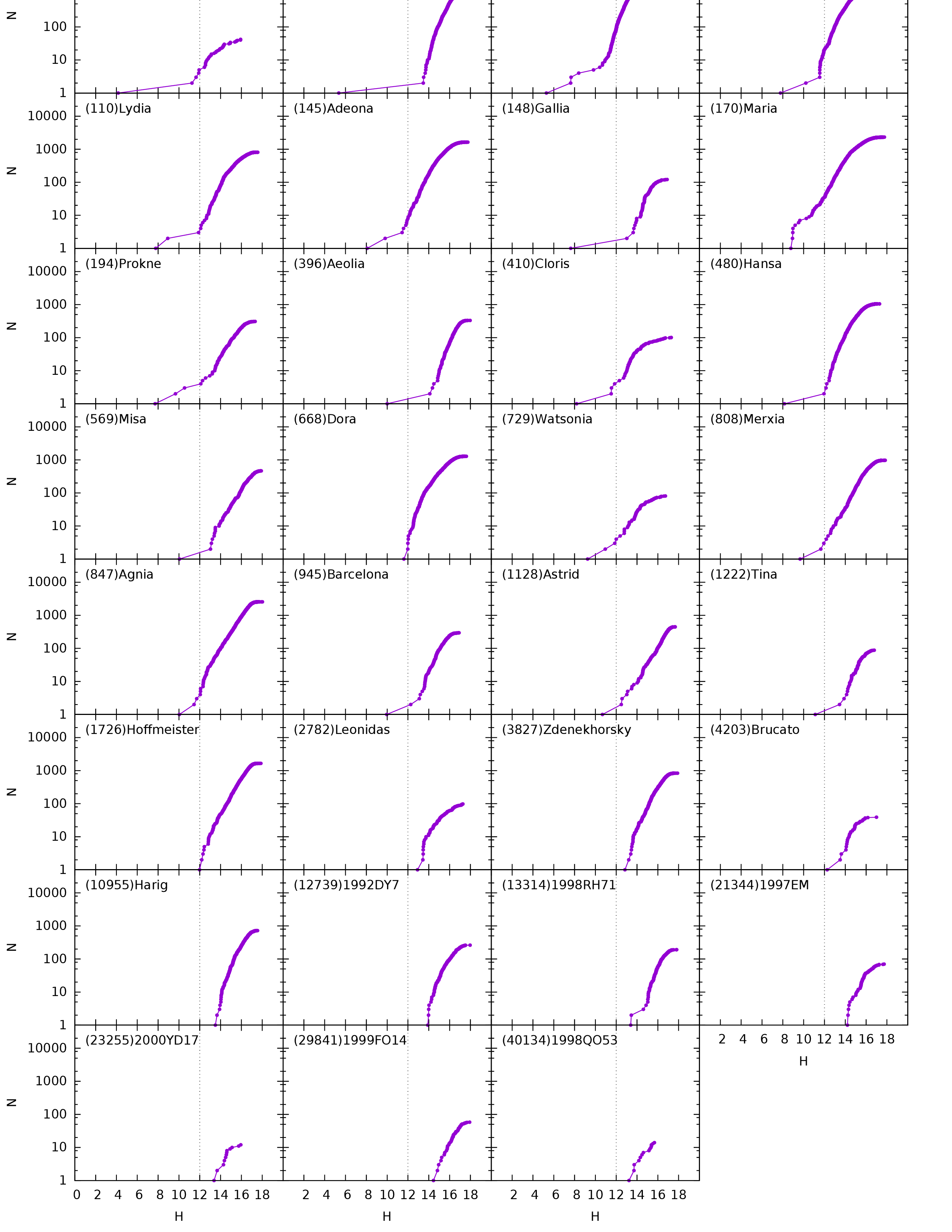} 
 \caption{Absolute magnitude distributions of the asteroid families in the middle belt.}
   \label{Fig:sfdmid}
\end{center}
\end{figure}

\begin{figure}[h!b]
\begin{center}
\includegraphics[width=\textwidth]{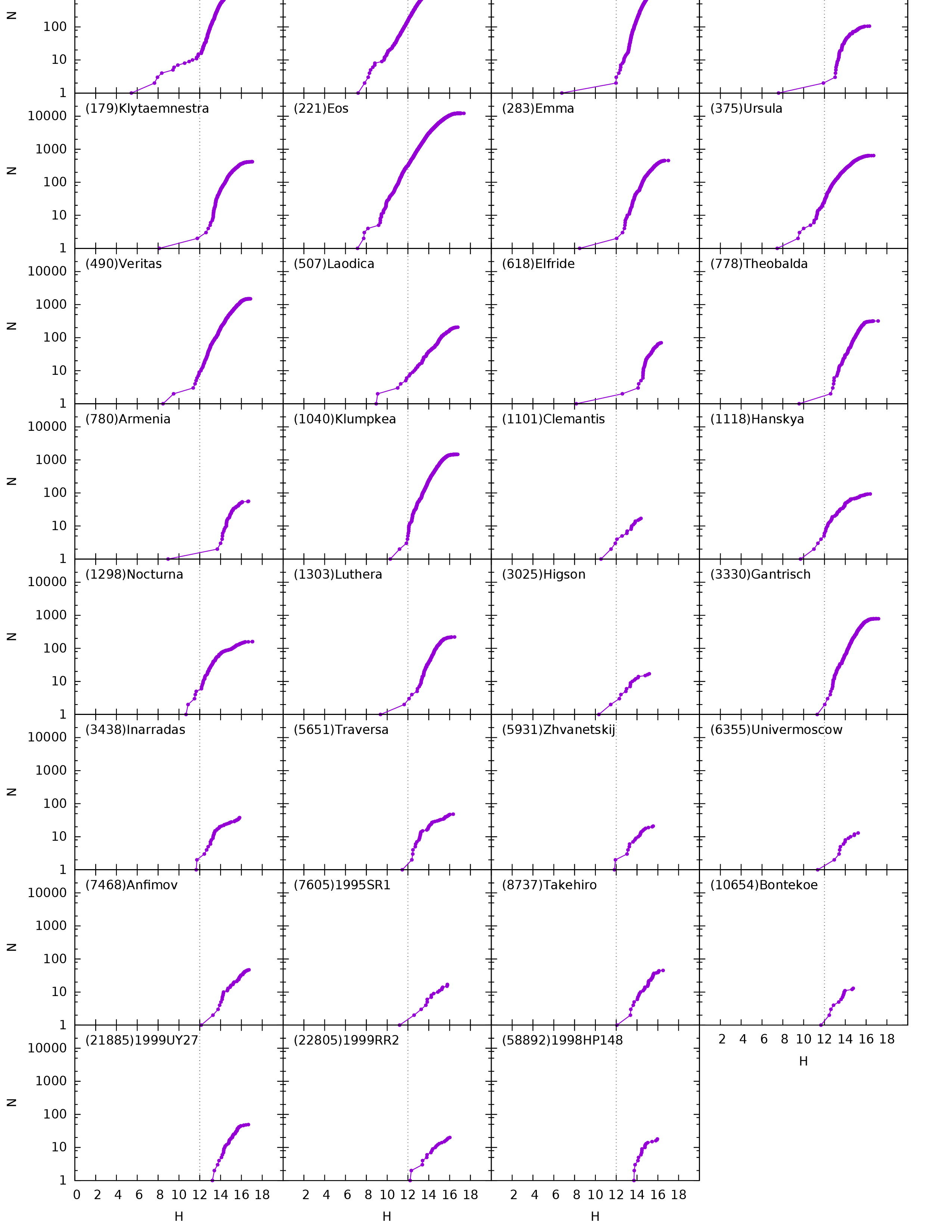} 
 \caption{Absolute magnitude distributions of the asteroid families in the outer belt.}
   \label{Fig:sfdout}
\end{center}
\end{figure}

Using the asteroid family classification of \cite{Milani2014,Milani2016} we produced the absolute magnitude distributions for each family in the middle (\autoref{Fig:sfdmid}) and outer belts (\autoref{Fig:sfdout}).   
There is a key difference in the size distributions of asteroid families in the two regions: In the middle belt, all the families with large parent bodies are cratering events, whereas all catastrophic families have parent bodies with $H>10$. On the contrary, in the outer belt there are two families with large parent bodies, those of (221) Eos and (24) Themis, which are of the catastrophic type. What this means is that in the middle belt there was no collisional event capable of producing a significant number of asteroids larger than magnitude H=12, as can be seen in \autoref{Fig:sfdmid}. Only the families of (10) Eunomia and (170) Maria have some small contribution of larger asteroids, as even these asteroids are smaller than magnitude H=10. On the other hand, the two aforementioned fragmentation families in the outer belt, have a substantial number of members in the ($9<H<12$) range of absolute magnitudes, as shown in \autoref{Fig:sfdout}, dominating in this way the composite size distributions of family members and consequently  the contaminated ``background'' population.

\begin{figure}[h!]
\begin{center}
\includegraphics[width=\textwidth]{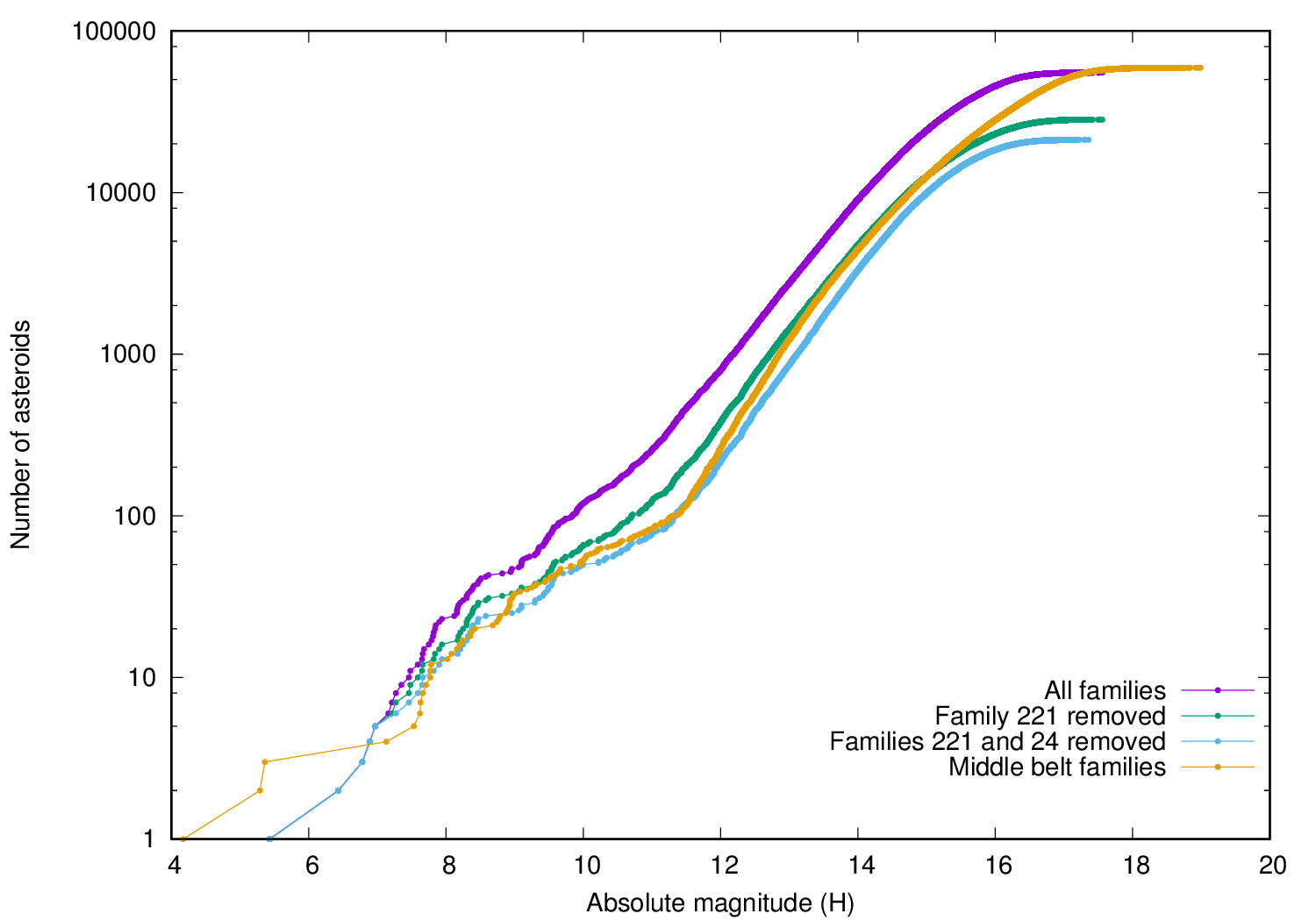} 
 \caption{Composite size distribution of the outer belt asteroid family members after removing the two big catastrophic type families Eos and Themis}
   \label{Fig:outnocat}
\end{center}
\end{figure}

By removing from the composite population of family members the ones originating from these two large catastrophic families, we can verify that they are the sole reason for the observed difference in the magnitude distributions between the middle and outer belt. Indeed, as shown in \autoref{Fig:outnocat}, the removal of the families of Eos and Themis gives a size distribution of family members with much shallower slope in the ($9<H<12$) range, resembling that of the middle belt.

Based on the above we can reach two important conclusions: The magnitude distribution of background asteroids in the pristine zone reflects qualitatively that of the primordial population of main-belt asteroids in the range $9<H<12$. This is based on the facts that in this range the asteroid families in the pristine zone have a completely different distribution from the background; Moreover it is similar to the magnitude distribution of the background population in the middle zone in the same H-range and the difference with that of the outer zone is fully understood by the contamination from the catastrophic large-parent body families of Eos and Themis. As a consequence of this, we can draw the second conclusion, that is: There are no large ancient families of the catastrophic type which are not yet identified in the middle or pristine zone. We cannot claim the same for the outer belt; The catastrophic families of Eos and Themis, as we have shown, are responsible for the contamination of the background population with $H<12$, but they might not be solely responsible. We cannot exclude the potential existence of another large unidentifiable family whose signature in the SFD has been overwritten by Eos and Themis.

\section{The size distribution of the primordial asteroid population}

Even though we are confident that the background population of the pristine zone is not contaminated significantly by rogue family members in the range $9<H<12$, we know that it still does not represent exactly the primordial population of asteroids. The reason for this is that the background population we observe today has undergone both collisional and dynamical evolution over the age of the Solar System, causing a number of asteroids to be removed from the region. As a consequence, the slope of the primordial population of asteroids must have been steeper than the current one. In order to constrain the primordial slope, we need to quantify the effects of the collisional and the dynamical evolution, and compensate for them.  

\begin{figure}[h!]
\begin{center}
\includegraphics[width=\textwidth]{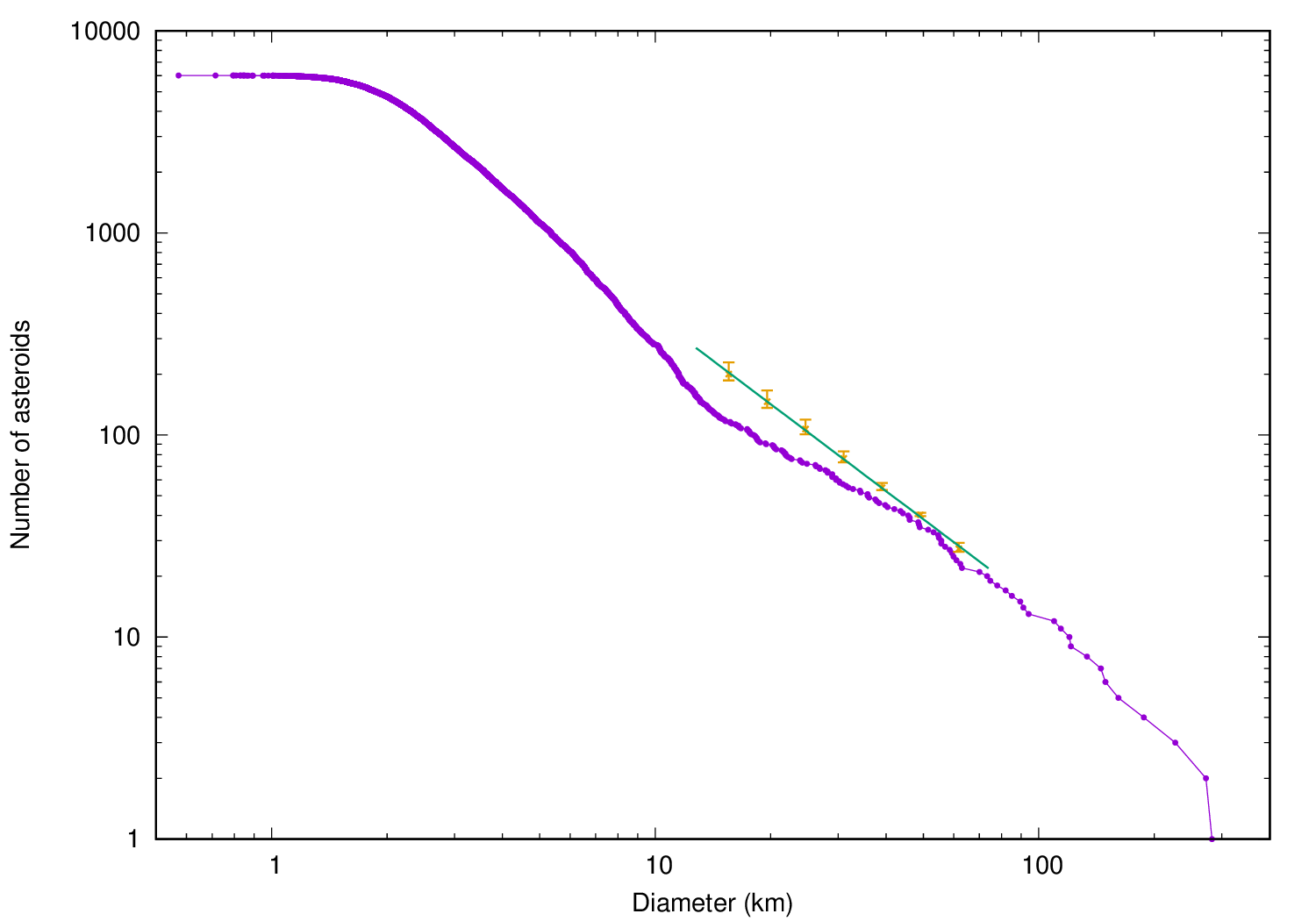} 
 \caption{Purple: The size distribution of the background population of the pristine zone, Orange:  The values at each bin after applying the corrections described in the text, and Green: The power law function fitting the corrected SFD in the  range $17.5<D(\rm{km})<70$, with a slope of $q_{\rm{c}}=-1.43^{+0.07}_{-0.05}$ }
   \label{Fig:sfd}
\end{center}
\end{figure}

We start by computing the cumulative size distribution from the magnitude distribution, making use of the formula $D(\rm{km})=\frac{1329}{\sqrt{p_{\rm{v}}}}10^{-0.2H}$ (\cite{Fowler1992}) and adopting a mean albedo of $(p_{\rm{v}}=0.092)$, as used in \cite{Bottke2005}. Doing this we obtain \autoref{Fig:sfd}, where the cumulative size distribution  of the background is shown in purple. The range  $9<H<12$ corresponds to approximately $17.5<D(\rm{km})<70$, and can be fit with a power law function with a slope $q\simeq-1$. Then, to compensate for the collisional erosion, we take into account the probability for an asteroid of a given size to have been catastrophically disrupted over the age of the Solar system. For this we need to use the collisional probabilities of asteroids in the pristine zone. Since the pristine zone appears to be a special region of the Main-Belt, it is not straightforward how to obtain these. One approach is to use the mean collisional probabilities of different diameter  main-belt asteroids taken from \cite{Bottke1994} (with updated collision frequencies kindly provided to us by the author). However, due to the fact that the number density of asteroids in the pristine zone is lower than the average of the main-belt, we expect the collisional probabilities therein to be different. Therefore, we need to calculate a new set of collisional probabilities for the pristine zone specifically.  Indeed, we calculated the collisional probabilities for target asteroids residing in the pristine zone, with the same sizes as those in \cite{Bottke1994}, and found them to be almost half as high. The two different sets of collisional probabilities will give different corrections to the slope of the size frequency distribution, as we will discuss later on.

\begin{table}[]
\centering
\caption{Sample absolute magnitudes and computed diameters and lifetimes.}
\label{Table:2}
\begin{tabular}{|c|c|c|}
\hline
\begin{tabular}[c]{@{}c@{}}Absolute \\ magnitude (H)\end{tabular} & Diameter (Km) & Lifetime (My) \\ \hline
13.25                                                             & 9.8092        & 4725.3         \\
12.75                                                             & 12.349        & 4995.9         \\
12.25                                                             & 15.547        & 5235.3         \\
11.75                                                             & 19.572        & 5693.0         \\
11.25                                                             & 24.64         & 6572.1         \\
10.75                                                             & 31.019        & 8266.1         \\
10.25                                                             & 39.051        & 10916.0        \\
9.75                                                              & 49.162        & 14984.9        \\
9.25                                                              & 61.892        & 20797.9        \\
8.75                                                              & 77.917        & 27593.9        \\
8.25                                                              & 98.092        & 33703.9        \\
7.75                                                              & 123.49        & 34905.0        \\ \hline
\end{tabular}
\end{table}

We are now able to  correct the slope of the cumulative size distribution, to better reflect the actual primordial size-distribution of planetesimals smaller than 70~km. To do so we use the following idea: The difference in lifetime of asteroid populations with different diameters leads to a difference in the rate at which these populations decay collisionally over time, and consequently the size distribution should be corrected accordingly.

For each diameter bin of \autoref{Table:2} we set up a simple Monte Carlo run of 100,000 test particles, that simulates the collisional decay over the age of the Solar system based on the respective lifetime. The result is a factor $f_{\rm{c}}(D)$ by which the observed population at each bin should be multiplied to compensate for the collisional grinding that has taken place.   

Another effect that has to be taken into account for the correction of the primordial size distribution of asteroids is the dynamical depletion. Over time, asteroids in the pristine zone drift secularly in semi-major axis due to the Yarkovsky effect, until they reach the powerful MMRs bounding the region, at which point they are ejected \footnote{We ignore here other dynamical effects, given that in the pristine zone there are no important resonances, capable of contributing significantly in the depletion of asteroids.} . This means that the initial population of asteroids in the pristine zone must have been larger than the current one. To compensate for this effect we devised another Monte Carlo scheme: For each diameter bin we create 10,000 fictitious asteroids with random initial conditions $(a,e,i)$ and random spin-axis obliquities $(\gamma)$ in the pristine zone. Then assuming a typical maximum drift speed of $(da/dt)_{max}=3\cdot10^{-4}\,\rm{AU/Myr}$ for an asteroid with D=1 km, each asteroid will drift over 4 billion years a distance: $4\cdot10^3 \times 3\cdot10^{-4} \times cos(\gamma)\times D^{-1}\,\rm{AU}$. We thus obtain the fraction of asteroids that have escaped the pristine zone and we can compute the corresponding correction due to the dynamical depletion $f_{\rm{d}}(D)$.  

Having obtained the corrections for both the collisional and the dynamical depletion of the primordial asteroid population in the pristine zone, we can compute the corrected size distribution. From the cumulative SFD of the pristine zone's background population we build the incremental size distribution, using the  bins in diameter from \autoref{Table:2}. Then we multiply the population in each bin by the corresponding factor $f(D)=1+f_{\rm{c}}+f_{\rm{d}}$, and compute the new corrected cumulative distribution as shown in \autoref{Fig:sfd} (upper ends of error bars). These points give a slope: $q_{\rm{high}}\simeq-1.50$. Using the collisional probabilities for target asteroids in the pristine zone, which are only half as high as those given in \autoref{Table:2}, the collisional lifetimes will be twice as long. Following the same procedure as before we find the second corrected SFD (lower ends of error bars) which has a slope: $q_{\rm{low}}\simeq-1.38$. The true collisional probabilities for each target diameter should be between these two values, and we select the arithmetic mean as the nominal ones, from which we obtain our final corrected slope of the primordial SFD:  $q_{\rm{c}}=-1.43^{+0.07}_{-0.05}$,  as shown in green in \autoref{Fig:sfd}.

Our computation, despite our efforts has some shortcomings that may affect the value of the slope we obtain for the primordial SFD. One shortcoming is that the removal of asteroid family members can never be perfect. Even in the pristine zone, where the families are few and well separated, there should be a small number of asteroids originating from collisions that are unidentifiable as family members by HCM. Still the value we obtained can be considered an upper bound to the slope of the primordial SFD.

\section{Conclusions}

In this work, we believe to have found evidence, by removing asteroid family members, that the primordial slope of the asteroids' SFD with $D<70\rm{km}$ was much shallower than the current one. This is in agreement with the predictions of \citet{Bottke2005}. We give, for the first time, an estimate of what that slope should have been, i.e. $q_{\rm{c}}=-1.43^{+0.07}_{-0.05}$. This is significantly shallower than the current slope of -1.8, which is also the slope predicted by streaming instability simulations. However, it is not clear to which size range the slope found in those simulations applies to. The fact that the  slope we measured below 70~km is shallower, suggests that the streaming instability slope (-1.8 for the cumulative distribution) applies for bodies larger than this threshold size, and that below 70~km the streaming instability process may be less efficient \citep{Klahr2016}.  Moreover, by comparing the SFD in the outer belt to those of the two other zones, we see exactly what \citet{Bottke2005} predicted: namely, that the SFD below the primordial ``knee'' (here at $D=70\rm{km}$) grew through catastrophic break-ups of the primordial asteroids with $D>70\rm{km}$. Here, we identified Eos and Themis to be the responsible for the increase of the SFD exponent in the outer zone. The fact that families contaminate substantially the background, steepens the asteroid SFD as a whole.

An interesting point arises in view of these results, that is which specific epoch in the evolution of the Main Belt, as part of the Solar System, corresponds to the designation ``primordial'' in the context of our work. Essentially the evolution of the population of large asteroids ($9<\rm{H}<12$ as discussed) should be size independent, given that the Yarkovsky effect is practically zero for these asteroids, and all other processes ( depletion, implantation, excitation) are indeed size independent. Therefore the remaining question regarding the exact definition of the  primordial SDF in terms of which era we are talking about, has to do with the last possible mixing of asteroids in semi-major axis, as this defines our zones. The Grand Tack \citep{walsh2011} is indeed the last large-scale process the Solar system suffered that resulted in a mixing of asteroids with respect to their semi-major axis. The giant planet instability \citep{tsiganis2005,levison2011} that happened after that is known to cause  mixing only in the eccentricities and inclinations of asteroids, but this does not change the population within each of the three zones as we use them. Therefore by primordial we refer to the post Grand Tack state of the Solar System which coincides with the time of the  depletion of the gas nebula.

Secondly, we found evidence that no catastrophic disruption of large ($D>70\rm{km}$) asteroids ever occurred in the middle and pristine zone. In fact, if this had happened in the primordial times, even if the corresponding family would have been dispersed in eccentricity and inclination, the imprint in the global SFD of the region would still be visible. This gives very important information on the asteroid belt collisional grinding in the old times. Let's assume here that the asteroid belt was substantially reshuffled in eccentricity and inclination about 4 Gyr ago, when the giant planets underwent a dynamical instability. Then, for large families like Eos and Themis, we can say that those that formed less than 4 Gyr ago are identifiable today and those that formed before are not, at least in the middle and pristine zones. We have 2 families formed in the last 4 Gyr in 3 out of 3 zones (Eos and Themis) and no comparable families formed before 4 Gyr ago in 2 out of 2 zones (for the outer belt we cannot exclude that there are no additional families).
Thus, the cumulative collisional evolution in the first 0.5 Gyr had to be less than that of the last 4 Gyr (for an equal cumulative collisional evolution we would expect ${2\over3} \cdot2$ families and we see none). This, again, is in perfect agreement with \citet{Bottke2005} and strongly suggests that the asteroid belt either was never massive or it was dynamically depleted very quickly \citep{Morbidelli2015}.  Of course, if the giant planet instability happened early (i.e. just after the removal of the gas from the disk, $\sim4.5$~ Gyr ago), this constraint becomes much less significant.

Finally we have shown that the designation ``pristine zone'' for the region of the main-belt with $2.82<a_{\rm{p}}<2.96$ is at least inaccurate. If fragments originating from the neighboring asteroid family of Eos can cross the 7/3 MMR with Jupiter and contaminate the region, it is safe to deduce that the ``barrier'' formed by this resonance is not completely impenetrable, but rather acts as an attenuator. Thus, not only the identified Eos members, but also background asteroids of the two regions can migrate due to the Yarkovsky effect across the resonance. If this is the case, the question why the ``pristine zone'' has a much lower number density of asteroids compared to the neighboring regions, remains open.  

\section*{Acknowledgments}

We would like to thank our colleague Miroslav Broz and the reviewers Alberto Cellino and Mikael Granvik for their constructive comments which helped improve the quality of this manuscript. This work was supported by the European Union [FP7/2007-2013], project: STARDUST-The Asteroid and Space Debris Network and has been conducted while G.Tsirvoulis was an intern at the Observatoire de la Côte d'Azur, Nice, France.

\bibliography{bib}

\end{document}